\DocumentMetadata{} %
\documentclass[sigconf,nonacm,natbib=true]{acmart}
\usepackage{booktabs} %
\usepackage{float}
\usepackage{enumitem}
\usepackage{multirow}
\usepackage{subfigure}
\usepackage{colortbl}
\usepackage{algorithm}
\usepackage{algpseudocode}  %
\usepackage{forest}
\usepackage{tikz}
\usepackage{mhchem}
\usetikzlibrary{positioning}

\usepackage{wrapfig}
\usepackage{caption}  %
\usepackage[most]{tcolorbox}

\theoremstyle{definition}

\newtheorem{discussion}{}
\newcommand{\code}[1]{{\small\ttfamily#1}}
\setcopyright{cc}

\begin{document}

\title{Syntax Meets Semantics: Understanding Scientific Formulae}
\renewcommand{\shorttitle}{Syntax Meets Semantics: Understanding Scientific Formulae}

\author{Yuni Susanti}
\affiliation{%
  \institution{FIZ Karlsruhe}
  \city{Berlin}
  \country{Germany}}
\email{yuni.susanti@fiz-karlsruhe.de}

\author{Moritz Schubotz}
\affiliation{%
  \institution{FIZ Karlsruhe}
  \city{Berlin}
  \country{Germany}}
\email{moritz.schubotz@fiz-karlsruhe.de}

\renewcommand{\shortauthors}{Y. Susanti et al.}

\begin{abstract} 
Scientific formulae are a fundamental component of scholarly communication, yet their dual nature -- as structured \textit{syntax} 
and carriers of \textit{semantics} -- remains underexplored in scholarly information retrieval. Although prior studies show that jointly modeling syntactic and semantic modalities improves retrieval performance, the relationship between their underlying representations has not been systematically investigated. In this work, we empirically study cross-modal correspondence between formula syntax and semantics. We find that their native representation spaces exhibit extremely weak observable correspondence despite strong latent correlation, indicating a substantial representation mismatch between the two modalities. We further evaluate whether this mismatch can be reduced using standard representation learning and alignment techniques. We represent syntactic structure using graph-based encoders and semantic information using text-based encoders, then apply contrastive learning to induce a shared representation space. Results show that the learned alignment substantially improves cross-modal retrieval, suggesting that explicit representation learning can recover correspondence absent from the original representation spaces.

\end{abstract}

\vspace{-2mm}

\vspace{-3mm}

\maketitle

\section{Introduction}
\label{sec:intro}
Scientific formulae play a central role in scientific communication, functioning both as precise symbolic expressions and as dense carriers of conceptual meaning. This duality manifests in two 
fundamentally different representations of formula. The first is its \textbf{syntax}, typically captured 
using markup languages such as LaTeX or MathML \cite{miller2001mathml,kohlhase2006omdoc,mathrep}. For example, the physics formula $F = m\cdot a$ has syntactic representation as an \textit{equal} ($=$) node whose left child is the symbol $F$ and 
right child is a \textit{multiplication} node with operands $m$ and $a$. 
The second is its \textbf{semantic}, which corresponds to the natural-language description or scientific concept expressed by the formula;
in the previous example, that \textit{force} equals \textit{mass} multiplied by \textit{acceleration}, defining Newton’s second law of motion. 
Similar examples exists for mathematical identities (e.g., $(a+b)^2=a^2+2ab+b^2$), and in chemical formulae such as $\ce{H_2O}$.
Although these two modalities jointly determine how humans read, and ultimately understand scientific formula, their relationship remains insufficiently understood from a computational perspective.

Many downstream tasks in scholarly IR implicitly rely on the assumption that syntactic similarity approximates semantic similarity. For instance, formula retrieval systems typically rely on tree-based representations~\cite{wang2021tree,wangrettree,zhong2019structural,davila2016tangent3,davila2017layout} or text- and substructure-based overlap~\cite{miller2003technical,misutka2008extending,kumar2012structure}, while embedding-based methods tokenize formulae directly, implicitly assuming that structural representations capture semantic information~\cite{mathbertjia,mathbertpeng,mathrep}. Although prior work has shown that jointly modeling syntactic and semantic information improves retrieval effectiveness\cite{li-2025-formula,mathbertjia,ssemb}, comparatively little is known about the intrinsic relationship between their underlying representations. In particular, it remains unclear to what extent these modalities correspond in their native representation spaces and whether any mismatch can be systematically characterized.

To address this question, we first conduct a quantitative analysis of the relationship between syntactic and semantic representations using alignment measures spanning instance-level, global, and latent-space correspondence. Our analysis reveals a discrepancy: the two modalities exhibit extremely weak observable correspondence (R@10 < 0.02), yet share strong latent correlations (Top-10 CCA = 0.91), indicating that related information is organized differently in their respective representation spaces. Motivated by this observation, we then investigate whether this mismatch can be reduced through learned cross-modal representations. Using graph-based encoders for formula syntax and text-based encoders for semantics, we evaluate standard contrastive learning as a mechanism for inducing a shared embedding space. The resulting representations substantially improve cross-modal retrieval (up to +0.52 Recall), suggesting that explicit representation learning can recover correspondence that is not directly accessible in the original representation spaces. Overall, this work provides a systematic empirical characterization of the relationship between syntactic and semantic representations of scientific formulae and examines how learned alignment can bridge the gap between them. Our dataset and code are publicly available for future research.\footnote{\url{https://github.com/susantiyuni/formula-alignment}}

\section{Related Work}
\label{relwork}

Research in formula-related information retrieval has traditionally focused on syntactic structure. Early approaches model formulae and their surrounding textual context as sequences of tokens, employing token-based representations for retrieval and matching \cite{eqembed,Topiceq,yuan2016mathematical}. Owing to the inherently hierarchical nature of formulae, many subsequent studies adopt tree-structured representations to explicitly capture their structure \cite{wang2021tree,wangrettree,zhong2019structural,davila2016tangent3,davila2017layout}. More recent work learns formula embeddings directly from LaTeX or MathML sequences, or from graph-structured operator trees. These approaches range from symbol-level embeddings~\cite{eqembed, Topiceq, MathAMR2022, Thanda2016, Tangent-CFT2019, Dai2020}, including \textit{symbol2vec} and \textit{formula2vec}~\cite{gao2017preliminaryexplorationformulaembedding}, to higher-level graph embeddings that encode global formula structure \cite{Song2021}.

Semantic information is typically captured via natural-language descriptions,
and transformer-based models, such as BERT \cite{devlin2019bert} and Sentence-BERT (SBERT) \cite{reimers-2019-sentence-bert}, have become standard tools for encoding scientific text. Several works \cite{mathbertjia,mathbertpeng,MathBERTa} extend this by pretraining the models on math-intensive corpora.
These models and their variants have been applied to a range of scientific tasks, including math-specific such as math question answering \cite{MathBERTa} and scientific document retrieval and recommendation~\cite{cohan-etal-2020-specter}.

\section{Formula Representation}
\label{sec:alignment}

Accurate representation of scientific formulae requires capturing both their syntactic and semantic dimension, as purely syntactic representations may miss conceptual meaning, while purely semantic representations may ignore precise structural constraints. 
To enable a systematic comparison between these two modalities, we construct separate syntactic and semantic representations for every formula. 
These representations serve as the basis for the experiments in Section~\ref{sec:eval}.

\subsection{Syntactic Representation}
\label{sec:syntax}
Scientific formulae can be interpreted as hierarchical symbolic structures in which operators and functions express relationships among identifiers and numerical constants. To capture these patterns, we represent each formula as an operator tree (OPT) derived from its \textit{content} MathML encoding. This convert a formula into a tree whose nodes correspond to operators, functions, and symbols, capturing semantic via its syntax~\cite{zhong2019structural,mathbertpeng,ssemb}.

\medskip
\noindent \textbf{Input Construction}. From a MathML expression, we construct a normalized OPT by assigning each element to one of three canonical node types: \textit{symbolic} (\texttt{mi}, \texttt{mn}, \texttt{mtext}), \textit{operator} (\texttt{mo}), and \textit{functional} nodes for structured constructs such as superscripts (\texttt{msup}), subscripts (\texttt{msub}), fractions (\texttt{mfrac}), and square roots (\texttt{msqrt}). 
Formally, for each MathML node $u$, we generate an OPT node
\[
\mathrm{OPT}(u) = (\mathrm{type}(u), \mathrm{value}(u), \mathrm{children}(u)),
\]
where $\mathrm{type}(u)$ indicates the canonical category, $\mathrm{value}(u)$ stores the associated token (e.g., ``$x$'', ``$+$'', ``power''), and $\mathrm{children}(u)$ is the ordered list of child nodes. Each child is annotated with a predefined semantic role (e.g., \textit{base} and \textit{exp} for superscripts, \textit{num} and \textit{den} for fractions), preserving both hierarchical and semantic relations of the formula in the symbolic OPT representation.

\medskip
\noindent\textbf{Graph-based Syntactic Encoder.}
Each OPT tree representing the syntactic form of formula (see \S\ref{sec:syntax}) is converted into a directed graph $G = (V, E)$.
for graph neural network (GNN) processing. For every node $v \in V$, we assign a canonical node type \{\textit{symbol}, \textit{operator}, \textit{function}\}, where each encoded as a learned embedding. For every edge $(i,j) \in E$ linking a parent node to one of its children, we attach a two-dimensional edge attribute:
\[
\mathrm{edge\_attr}(i,j)
  = (\mathrm{role}_{ij}, \mathrm{pos}_{ij}),
\]
where $\mathrm{role}_{ij}$ is an integer encoding the semantic role 
(e.g., \textit{base}, \textit{exp}, \textit{num}, \textit{den}, \textit{sub}, \textit{sup}, \textit{arg}), 
and $\mathrm{pos}_{ij} \in \mathbb{N}$ denotes the child position index.
This design 
allows the model to encode distinctions between common structural relations in scientific formulae, such as \textit{numerator} vs. \textit{denominator} or \textit{base} vs. \textit{exponent}.

To encode the structural information present in $G$, we use 
a GINE~\cite{hu2020strategies}-based message-passing network with node-type embeddings and role- and position-aware edge features to encode formula structure.
Each node embedding is initialized as
\[
h_i^{(0)} 
  = \mathrm{emb}_{\mathrm{label}}(x_i^{\mathrm{label}})
    + \mathrm{emb}_{\mathrm{type}}(x_i^{\mathrm{type}}).
\]

Each edge $(i,j)$ receives an embedding
\[
e_{ij}
  = \mathrm{emb}_{\mathrm{role}}(\mathrm{role}_{ij})
    + \mathrm{emb}_{\mathrm{pos}}(\mathrm{pos}_{ij}).
\]

We further apply two stacked GINEConv~\cite{hu2020strategies} layers 
followed by global mean pooling to obtain final 
syntactic representation
\[
x_{\mathrm{struct}} 
  = W_{\mathrm{proj}} 
    \cdot \mathrm{pool}(h_i^{(L)}).
\]

\subsection{Semantic Representation}
\label{sec:semantic}
We compute semantic representation for each formula by leveraging textual descriptions of the underlying scientific concepts it expresses. The intuition is that a formula reflects domain knowledge through its associated concepts, often documented in external knowledge sources. Embedding these descriptions provides a high-level semantic signal that is not captured by symbolic structure. 

\medskip
\noindent \textbf{Input Construction}. To construct a comprehensive semantic representation, all available textual annotations, (e.g., labels, descriptions) are aggregated into a unified text sequence. Duplicate information is removed to preserve only unique semantic content while maintaining the contextual richness of the representation. Finally, for a formula $i$, $D_i = \{ d_{i1}, \dots, d_{ik} \}$ denote the final set of distinct textual annotations associated with formula $i$. 

\medskip
\noindent\textbf{Transformer-based Semantic Text Encoder.} The resulting text sequence is subsequently encoded into dense vector embeddings using a pretrained language model (e.g., \textit{SentenceTransformer}~\cite{reimers-2019-sentence-bert}):

\[
e_{ij} = f_{\mathrm{encoder}}(d_{ij}) \in \mathbb{R}^{d_2}.
\]

To handle multiple descriptions per formula, embeddings are aggregated.
This yields a fixed-dimensional semantic embedding $X_{\mathrm{sem}}^{(i)} \in \mathbb{R}^{d_2}$ representing the combined semantic meaning of the formula, capturing contributions from each concept's description as well as the overall formula description, if exists.

\begin{figure}[h]
    \centering
    \includegraphics[width=0.4\textwidth]{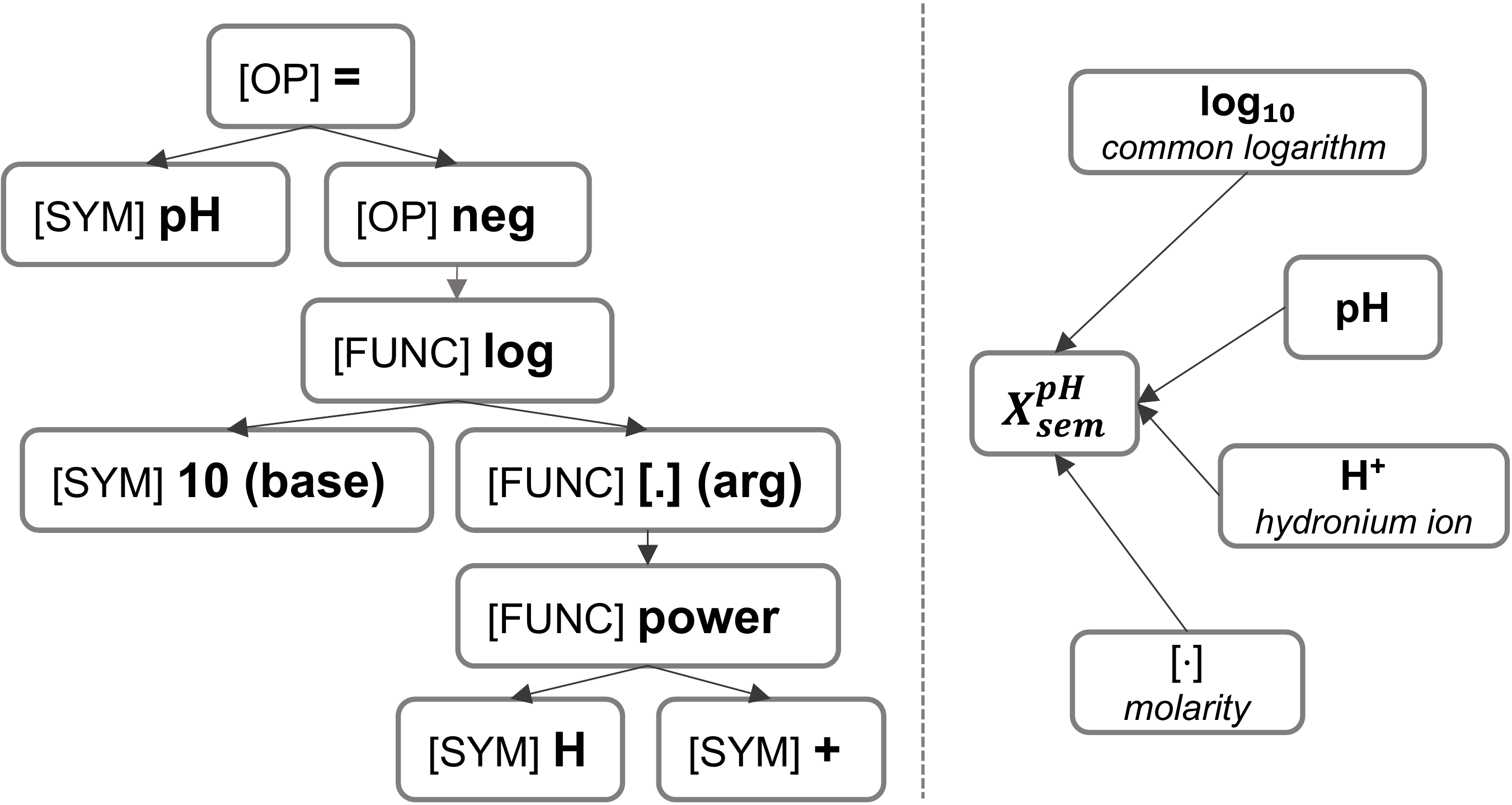}
    \caption{Formula $\text{pH} = -\log_{10}[H^+]$ with its \textit{OPT}-based syntactic (a) and \textit{concept}-based semantic (b) representation.}
    \label{fig:ill}
\end{figure}

As an example, consider the formula 
$
\text{pH} = -\log_{10}[H^+],
$
linking the concept \textit{pH} to the concepts \textit{hydronium ion} (\code{H$^+$)}, \textit{molarity} (\code{[.]}), and \textit{common logarithm} (\code{log}). Its semantic representation is,

\[
X_\mathrm{sem}^{(\text{pH})} = 
e_\text{pH} + e_\text{H$^+$} + e_\text{log} + e_\text{[.]}
\]

where $e_\text{pH}, e_\text{H$^+$}, e_\text{log}$ are the embeddings of each of the linked concepts. This aggregation ensures that the formula embedding reflects its multi-concept structure and the domain-specific interpretations of each concept. Figure~\ref{fig:ill} illustrates the syntactic and semantic representation of the $pH$ formula. 

\section{Evaluation Experiments}
\label{sec:eval}
Having obtained structural and semantic representations for each formula, we next investigate the extent to which these modalities \textit{naturally} correspond before any learned alignment is applied (\S\ref{sec:natural}: \textit{Quantifying Syntax--Semantic Correspondence}). 
This is followed by the second experiment (\S\ref{sec:jointemb}: \textit{Learning Syntax--Semantic Alignment}), investigating whether standard cross-modal alignment techniques can effectively learn shared space between the two modalities.
\subsection{Dataset}
\label{sec:dataset}
We construct a \textit{gold} dataset of 2,405 scientific formulae paired with their associated scientific concepts and descriptions from Wikidata, using formula-bearing properties (\code{P2534}:\textit{defining formula} and \code{P7235}:\textit{in defining formula}). We use Wikidata because it provides structured, high-quality formula--concept links, making it ideal for studying syntax--semantic correspondence and alignment. We publish the full dataset on our repository (see Page~1).

\subsection{Quantifying Syntax--Semantic Correspondence}
\label{sec:natural}
We quantify the syntax--semantic correspondence of scientific formulae using the following measures:
\begin{discussion}{\textbf{Instance-Level Correspondence.}}
To quantify \textit{instance}-level correspondence between the two modalities, we compute \textit{cosine similarity} between each formula syntactic and semantic representation. We also perform \textit{cross-modal retrieval}, measuring whether the correct semantic embedding can be retrieved from a structural query representation. Additionally, we evaluate geometric compatibility using \textit{Orthogonal Procrustes alignment}~\cite{procschonemann1966generalized}, which finds the optimal rigid transformation from the syntactic to the semantic space. These measures capture how well the two spaces align \textit{without} any learned transformation.
\end{discussion}
\vspace{-2mm}
\begin{discussion}{\textbf{Global Representational Similarity.}}
To assess whether both modalities encode similar overall relationships, we compute \textit{Representational Similarity Analysis} (RSA)~\cite{rsakriegeskorte2008representational} and \textit{Centered Kernel Alignment} (CKA)~\cite{ckakornblith2019similarity}. RSA compares the pairwise similarity structures induced by the two representations, while CKA quantifies alignment invariant to linear transformations and scaling. These metrics provide a global view of representational correspondence beyond individual formula pairs.
\end{discussion}
\vspace{-2mm}
\begin{discussion}{\textbf{Shared Latent Structure.}}
To evaluate whether the modalities share underlying latent factors, we apply \textit{Canonical Correlation Analysis} (CCA)~\cite{hotelling1936relations}. CCA finds linear projections that maximize correlation between the syntactic and semantic spaces, revealing if a shared latent space exists even when the original representation are misaligned. High CCA indicates both representations encode overlapping information in a linearly recoverable form.
\end{discussion}

\begin{figure}[t]
    \centering
        \includegraphics[width=0.5\textwidth]{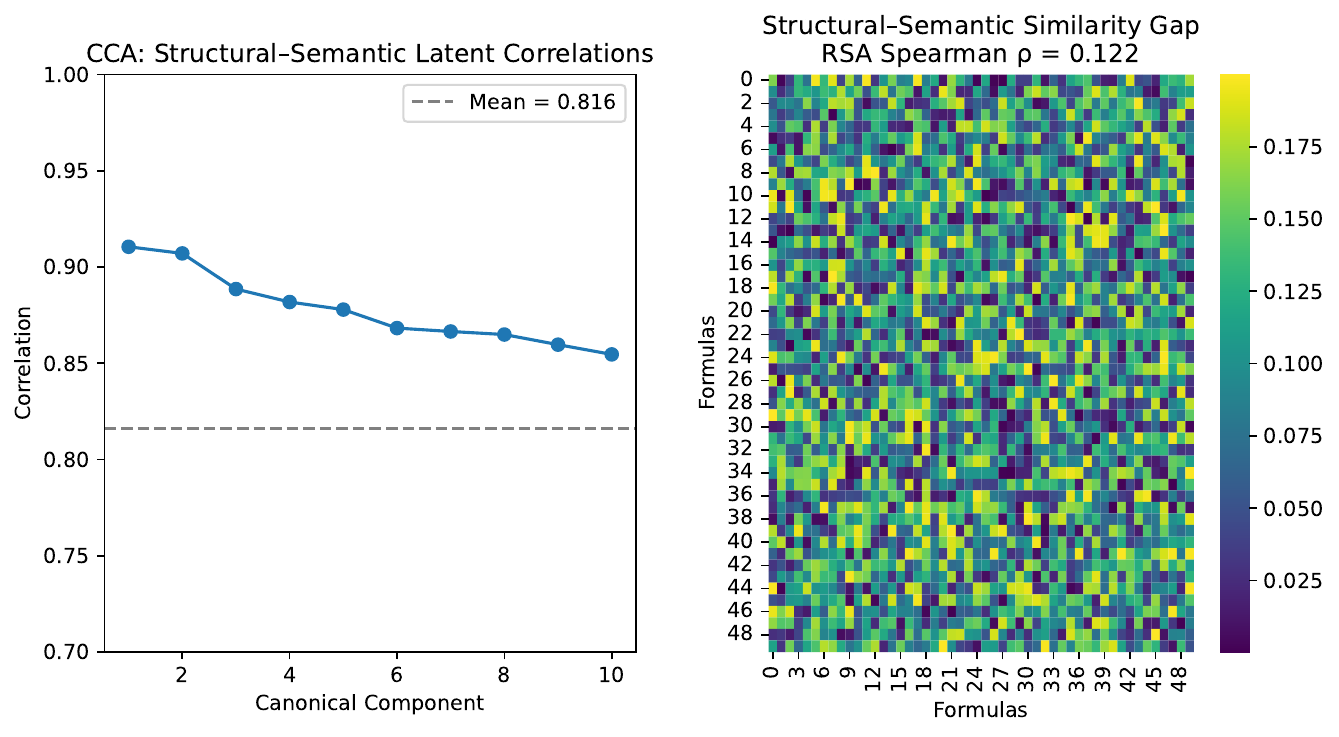}
    \caption{\textit{Raw} alignment analysis: CCA latent correlations (\textit{left}); syntax–semantic similarity gap (\textit{right}).}
    \label{fig:alignment}
\end{figure}
\medskip
\subsubsection{Results and Discussion}
Table~\ref{tab:alignment_metrics} summarizes the alignment analysis results, with selected results shown in Fig.~\ref{fig:alignment}.
We summarize the key findings, as follows.

\setlength{\abovecaptionskip}{3pt}  %
\setlength{\belowcaptionskip}{1pt}  %
\setlength{\columnsep}{7pt}   %
\begin{wraptable}[14]{r}{3.6cm}  %
\renewcommand{\arraystretch}{0.84}  %
\vspace*{-1.8\baselineskip}    %
\centering
\caption{\textit{Raw} alignment analysis metrics results.}
\label{tab:alignment_metrics}
\begin{tabular}{ll}
\toprule
\textbf{Metric} & \textbf{Value} \\
\midrule
Mean Cosine & .0847 \\
Median Cosine & .0939 \\
Std Cosine & .2114 \\
RSA Spearman $\rho$ & .1215 \\
Procrustes Error & .8299 \\
Linear CKA & .1566 \\
Top-10 CCA & .85--.91 \\
Mean CCA & .8161 \\
Recall@1 & .0006 \\
Recall@5 & .0068 \\
Recall@10 & .0125 \\
Permutation Test & .0847 \\
\bottomrule
\end{tabular}
\end{wraptable}

\vspace{1.5mm}
\noindent\textit{Weak direct alignment.} The mean cosine similarity between paired representation is 0.085 (median 0.094), indicating limited direct correspondence at the individual formula level. Cross-modal retrieval further confirms this mismatch with low accuracies (R@10 $<0.02$).
Similarly, orthogonal Procrustes alignment yields a high alignment error (0.83), suggesting that the two representations are not geometrically compatible under rigid transformations.
Nevertheless, permutation test shows that cosine alignment is significantly above chance (0.08), implying the presence of weak but non-random correspondence.

\vspace{1.5mm}
\noindent\textit{Modest global similarity.} Despite poor direct, instance-level alignment, global similarity analyses reveal moderate correspondence between the two representations. RSA yields a Spearman correlation of $\rho = 0.12$ (p < 0.001), indicating that formulae that are syntactically similar tend to be semantically similar more often than chance. Consistently, linear CKA reports a score of approximately 0.16, reflecting partial representational overlap that is invariant to scaling and rotation. These suggest that structural encoders capture aspects of semantic,
albeit in a distributed and indirect manner.

\vspace{1.5mm}
\noindent\textit{Strong shared latent structure.} 
In contrast to direct and global alignment metrics, CCA reveals substantial shared latent structure between the two modalities. The mean canonical correlation across components is 0.82, with the top canonical dimensions exceeding 0.90. This pronounced latent correspondence indicates that semantic information is strongly embedded within structural representations, but expressed in a transformed subspace that is not directly accessible through raw similarity measures.

\medskip
The weak direct alignment indicates that syntactic and semantic representations, despite describing related formulae, organize information differently in their native embedding spaces. This result does not imply that the two modalities are unrelated; rather, the strong latent correlation suggests that shared information exists but is expressed through different geometric structures.
This motivates the second half of the paper, where we investigate whether standard cross-modal alignment techniques can recover this hidden correspondence. Importantly, our goal is \textbf{\underline{not}} to propose a novel joint embedding method, but to empirically examine the extent to which existing representation alignment approaches can reconcile the geometric mismatch between formula syntax and semantics.

\subsection{Learning Syntax--Semantic Alignment}
\label{sec:jointemb}
The correspondence analysis in the previous section indicates that syntactic and semantic representations exhibit weak observable alignment despite sharing substantial latent structure. We therefore investigate whether this mismatch can be reduced through learned cross-modal representation alignment. In cross-modal alignment, the goal is to learn a shared space such that different modalities are integrated coherently. The joint representation of formula syntax and semantic should then jointly encode both information while maintaining its coherence, i.e., the similarity in either modality is meaningfully reflected in the space. 

\textit{Contrastive learning} (CL) has proven effective for aligning heterogeneous modalities, including image--text (e.g., CLIP)~\cite{radford2021clip} and code--documentation pairs \cite{feng2020codebert}. These methods learn shared embedding spaces where semantically corresponding items are brought closer together. Inspired by this line of work, we investigate \textit{contrastive alignment framework} for aligning syntactic and semantic representations of formulae.

\medskip
\noindent\textbf{Contrastive Cross-Modal Alignment.}
Given the structural embedding $x_{\mathrm{struct}}$ and semantic embedding
$x_{\mathrm{sem}}$ introduced in Section~\ref{sec:alignment}, we learn
modality-specific projection functions,
\[
z_{\mathrm{struct}}
=
W_{\mathrm{struct}}\,x_{\mathrm{struct}},
\qquad
z_{\mathrm{sem}}
=
W_{\mathrm{sem}}\,x_{\mathrm{sem}},
\]
which map both modalities into a shared latent space. The projection heads are
optimized using a \textit{symmetric InfoNCE objective}, following the contrastive
learning framework popularized by CLIP~\cite{radford2021clip}. The objective
encourages representations of matched syntax--semantic pairs to be close in the
shared embedding space while pushing unmatched pairs farther apart.

Formally, for a batch of
paired embeddings $\{(z_i^{\mathrm{struct}}, z_i^{\mathrm{sem}})\}_{i=1}^{B}$, we employ the loss function 
as follow:
\[
\mathcal{L}_{\mathrm{cross}}
=
\frac{1}{2} \left[
  \mathrm{CE}\!\left(
    \frac{z^{\mathrm{struct}} (z^{\mathrm{sem}})^\top}{\tau}
  \right)
  +
  \mathrm{CE}\!\left(
    \frac{z^{\mathrm{sem}} (z^{\mathrm{struct}})^\top}{\tau}
  \right)
\right],
\]
where $\tau$ denotes contrastive temperature. During training, both modalities are $\ell_2$-normalized. To support consistency within each modality, we apply additional \textit{intra-modal} losses:
\[
\mathcal{L}_{\mathrm{struct}}
  = \mathrm{InfoNCE}(z^{\mathrm{struct}}, z^{\mathrm{struct}}),
\quad
\mathcal{L}_{\mathrm{sem}}
  = \mathrm{InfoNCE}(z^{\mathrm{sem}}, z^{\mathrm{sem}}).
\]

\medskip
\subsubsection{Experiment Settings}
We evaluate the learned alignment on \textit{cross-modal} retrieval task, considering two retrieval directions: 
\begin{enumerate}[leftmargin=0.5cm]
    \item \textbf{Syntactic → Semantic}, where the model retrieves the correct semantic representation given a formula’s syntactic form,
    \item \textbf{Semantic → Syntactic}, where the model retrieves the correct syntactic representation given its semantic representation.
\end{enumerate}
We evaluate the retrieval performance using \textit{Recall@k} ($k=1,5,10$) and \textit{Mean Reciprocal Rank} (MRR) metrics.
For comparison, we consider the following retrieval-based alignment approaches:

\begin{enumerate}[leftmargin=0.5cm]
    \item[(a)] \textbf{BM25}, a non-neural retrieval model where each formula is linearized into a token sequence and all associated textual information are aggregated into a single corpus against which BM25 scores are computed.
    \item[(b)] \textbf{Dual Encoder (CL)}: both formula syntax 
    and semantic 
    are encoded with the same transformer model and trained via constrastive learning (see \S\ref{sec:jointemb}), following general design of prior work on formula retrieval~\cite{li-2025-formula}. We implement the contrastive learning on two transformer models:
    \begin{itemize}
        \item \textbf{SBERT} (\code{all-mpnet-base-v2}),
        \item \textbf{MathBERT}~\cite{mathbertjia}.
    \end{itemize}
    For each, variants \textit{without} any contrastive learning {\textcolor{blue}{(\textit{w/o} CL)}} are also included.
    \item[(c)] \textbf{Graph--Text Encoder CL}, which applies contrastive learning directly to align graph-based syntactic representations with transformer-based semantic text encoder; also in prior work\cite{ssemb}.

\end{enumerate}

In all experiments, we use the dataset described in \S\ref{sec:dataset}. To account for its relatively small size, we evaluate with 5-fold cross-validation (0.8/0.2 train–test split) to ensure robust evaluation. 
We release all code and model training hyperparameters on GitHub.

\medskip
\subsubsection{Results and Discussion}
Table~\ref{tab:result} summarizes the results of all experiments. BM25, as a \textit{non-neural} baseline, performs poorly on both retrieval directions (R@10 $<$ 0.07, MRR $<$ 0.05).
This is expected, as BM25 relies on exact token overlap and treats formulae as \textit{bags-of-words}, ignoring syntactic and semantic cues. In our dataset, there is minimal lexical overlap between syntactic tokens and textual descriptions (e.g., {\small{$\sqrt{2}$}} vs. \textit{square root of 2}), limiting BM25’s ability to identify correct pairs. Using a pretrained neural embedding model (SBERT/MathBERT) \textit{without} contrastive learning (\textit{w/o} CL) improves performance over BM25, though the gain remains modest (R@10 $<$ 0.28, MRR $<$ 0.15).

In contrast, neural-based contrastive alignment 
(both Dual Encoders and Graph--Text CL) 
significantly outperform BM25 (up to +0.52 R@10) and neural baseline without CL (up to +0.47 R@10); all statistically significant ($p < 0.005$).
These results confirm 
that the observed mismatch between syntactic and semantic representations can be effectively reduced through learned shared embedding spaces. Importantly, the improvements are consistent across model architectures, suggesting that the gain primarily stems from explicit alignment rather than specific encoder design choices.

From a structural perspective, graph-based encoders provide the strongest performance. In graph-based encoding, node types and role- and position-aware edges preserve hierarchical and relational structure of formula components.
This suggest that explicitly modeling hierarchical formula structure is beneficial for cross-modal alignment. 
Nevertheless, the transformer-based dual encoders with SBERT/MathBERT also perform competitively, suggesting that sequential representations already capture a substantial portion of the syntactic information relevant for semantic matching.

Notably, performance remains largely symmetric across retrieval directions, indicating that the learned representations support bidirectional cross-modal retrieval. This is consistent with our earlier observation that syntactic and semantic representations share strong latent structure despite weak observable correspondence. Overall, these results suggest that explicit representation learning can recover meaningful alignment between the two modalities, even when such correspondence is not directly present in the original embedding spaces.

\begin{table}[]
\small
\renewcommand{\arraystretch}{0.80}  %
\centering
\caption{Cross-modal retrieval results and ablation \underline{without} contrastive learning {\textcolor{blue}{(\textit{w/o} CL)}}, averaged over 5-fold CV.}
\label{tab:result}
\begin{tabular}{@{}lcccc@{}}
\toprule
\multicolumn{1}{l|}{methods} & \multicolumn{1}{c}{R@1} & R@5 & R@10 & MRR \\ 
\midrule
\rowcolor[HTML]{FFCCC9}\multicolumn{5}{c}{\textbf{Syntactic $\rightarrow$ Semantic}} \\
\midrule
\multicolumn{1}{l|}{BM25} 
& .011 & .030 & .052 & .033 \\ 
\multicolumn{1}{l|}{\quad (\textit{std ±})} 
& \footnotesize{(.003)} & \footnotesize{(.005)} & \footnotesize{(.010)} & \footnotesize{(.003)} \\ 

\multicolumn{1}{l|}{SBERT~\cite{reimers-2019-sentence-bert} Dual Encoder \textcolor{blue}{(\textit{w/o} CL)}} 
& .020 & .061 & .100 & .052 \\ 
\multicolumn{1}{l|}{\quad (\textit{std ±})} 
& \footnotesize{(.008)} & \footnotesize{(.005)} & \footnotesize{(.009)} & \footnotesize{(.006)} \\ 

\multicolumn{1}{l|}{MathBERT\cite{mathbertjia} \textcolor{blue}{(\textit{w/o} CL)}} 
& .015 & .051 & .090 & .046 \\ 
\multicolumn{1}{l|}{\quad (\textit{std ±})} 
& \footnotesize{(.003)} & \footnotesize{(.004)} & \footnotesize{(.013)} & \footnotesize{(.003)} \\ 

\midrule
\multicolumn{1}{l|}{SBERT~\cite{reimers-2019-sentence-bert} Dual Encoder (CL)} 
& .224 & \textbf{.450} & .556 & .334 \\ 
\multicolumn{1}{l|}{\quad (\textit{std ±})} 
& \footnotesize{(.016)} & \footnotesize{(.028)} & \footnotesize{(.028)} & \footnotesize{(.015)} \\ 

\multicolumn{1}{l|}{MathBERT\cite{mathbertjia} (CL)} 
& .185 & .360 & .447 & .274 \\ 
\multicolumn{1}{l|}{\quad (\textit{std ±})} 
& \footnotesize{(.005)} & \footnotesize{(.015)} & \footnotesize{(.020)} & \footnotesize{(.007)} \\ 

\multicolumn{1}{l|}{Graph--Text Encoder CL} 
& \textbf{.238} & .444 & \textbf{.574} & \textbf{.345} \\ 
\multicolumn{1}{l|}{\quad (\textit{std ±})} 
& \footnotesize{(.014)} & \footnotesize{(.014)} & \footnotesize{(.014)} & \footnotesize{(.011)} \\ 

\midrule
\rowcolor[HTML]{FFCCC9}\multicolumn{5}{c}{\textbf{Semantic $\rightarrow$ Syntactic}} \\
\midrule
\multicolumn{1}{l|}{BM25} 
& .017 & .047 & .072 & .041 \\ 
\multicolumn{1}{l|}{\quad (\textit{std ±})} 
& \footnotesize{(.006)} & \footnotesize{(.009)} & \footnotesize{(.008)} & \footnotesize{(.006)} \\ 

\multicolumn{1}{l|}{SBERT~\cite{reimers-2019-sentence-bert} Dual Encoder \textcolor{blue}{(\textit{w/o} CL)}} 
& .082 & .201 & .280 & .149 \\ 
\multicolumn{1}{l|}{\quad (\textit{std ±})} 
& \footnotesize{(.016)} & \footnotesize{(.028)} & \footnotesize{(.028)} & \footnotesize{(.015)} \\ 

\multicolumn{1}{l|}{MathBERT\cite{mathbertjia} \textcolor{blue}{(\textit{w/o} CL)}} 
& .038 & .092 & .141 & .075 \\ 
\multicolumn{1}{l|}{\quad (\textit{std ±})} 
& \footnotesize{(.011)} & \footnotesize{(.014)} & \footnotesize{(.018)} & \footnotesize{(.012)} \\ 

\midrule
\multicolumn{1}{l|}{SBERT~\cite{reimers-2019-sentence-bert} Dual Encoder (CL)}
& .230 & .454 & .557 & .340 \\ 
\multicolumn{1}{l|}{\quad (\textit{std ±})} 
& \footnotesize{(.010)} & \footnotesize{(.014)} & \footnotesize{(.012)} & \footnotesize{(.009)} \\ 

\multicolumn{1}{l|}{MathBERT\cite{mathbertjia} (CL)}
& .179 & .361 & .443 & .270 \\ 
\multicolumn{1}{l|}{\quad (\textit{std ±})} 
& \footnotesize{(.006)} & \footnotesize{(.012)} & \footnotesize{(.011)} & \footnotesize{(.007)} \\  

\multicolumn{1}{l|}{Graph--Text Encoder CL}
& \textbf{.250} & \textbf{.458} & \textbf{.580} & \textbf{.355} \\ 
\multicolumn{1}{l|}{\quad (\textit{std ±})} 
& \footnotesize{(.019)} & \footnotesize{(.019)} & \footnotesize{(.017)} & \footnotesize{(.018)} \\ 

\bottomrule
\end{tabular}
\end{table}

\section{Conclusion}
\label{sec:conclusion}

This work investigates the relationship between syntactic and semantic representations of scientific formulae from a cross-modal perspective. Through a multi-level analysis of natural correspondence, we show that the two modalities exhibit weak observable alignment in their native embedding spaces, despite sharing substantial latent structure. These findings reveal a geometric mismatch between symbolic and semantic representations: related information is encoded across both modalities but organized differently.

We further examine whether this mismatch can be reduced through learned representation alignment. Using graph-based structural encoders and text-based semantic encoders, we demonstrate that standard contrastive alignment techniques can substantially improve cross-modal retrieval performance. These results suggest that the correspondence between formula syntax and semantics is not directly accessible from independent embedding spaces, but can be recovered through explicit alignment objectives.

Overall, this work provides an empirical analysis of the relationship between formula modalities and a foundation for studying their integration in scholarly information systems. Future work will investigate larger and more diverse scientific corpora, as
well as exploring its application to downstream tasks such as neuro-symbolic reasoning and scientific knowledge discovery.

\begin{acks}
This research was supported by the LUMEN project, funded by the European Union's Horizon Europe Research and Innovation Programme (Grant Agreement No.~\href{https://cordis.europa.eu/project/id/101187940}{101187940}).
Views and opinions expressed are those of the authors only and do not necessarily reflect those of the European Union or the granting authority. 
\end{acks}

\bibliographystyle{ACM-Reference-Format}
\balance
\bibliography{ref} 
\clearpage

\end{document}